\RequirePackage{fix-cm}
\documentclass[smallextended]{svjour3}       
\smartqed  
\usepackage{graphicx}
\usepackage{amssymb}
\usepackage{bm}
\begin{document}

\title{Thin $^4$He films on alkali substrates: where do $^3$He atoms bind?}
\author{Massimo Boninsegni}


\institute{Department of Physics, University of Alberta, Edmonton, Alberta, Canada, T6G 2H5\\
              \email{m.boninsegni@ualberta.ca}
}

\date{Received: date / Accepted: date}

\maketitle

\begin{abstract}
The possible occurrence of bound states of $^3$He atoms in the vicinity of a weakly attractive substrate coated with a thin superfluid $^4$He film is investigated by first principle computer simulations. No evidence is seen of such bound states, even in the case of the weakest substrate, i.e., Cs; a single $^3$He atom always binds to the free $^4$He surface, regardless of the thickness  of the $^4$He film. A comparison of $^4$He density profiles computed in this work with those yielded by the Density Functional approach that led to the prediction of $^3$He bound states  near the substrate, shows that the latter may not have afforded a sufficiently accurate structural description of the adsorbed $^4$He film.
\end{abstract}

\section{Introduction}
\label{intro}
The existence of a bound state of a $^3$He atom at a free liquid $^4$He surface was first proposed by Andreev \cite{andreev1966}, as an essentially {\em ad hoc} hypothesis to account for the observed behaviour of the surface tension of isotopic helium mixtures, significantly different from that of bulk $^4$He \cite{atkins1965}. The variational theory of Lekner \cite{lekner1970} provided a qualitative and semi-quantitative microscopic explanation for the occurrence of such bond state, tying it to the monotonic decrease of the local $^4$He density on approaching the surface.
\hfil\break\indent
Subsequently, Pavloff and Treiner \cite{pavloff1991,treiner1993} proposed that a similar physical mechanism could also underlie a $^3$He bound state at the interface of superfluid $^4$He and a weakly attractive (e.g., alkali) substrate, on which no solid $^4$He layer forms. Using a ground state density functional (DFT) approach, they computed the binding energy of a $^3$He bound state at the interface between superfluid $^4$He and different alkali substrates, obtaining a value close to 4 K for all of them \cite{treiner1993}. 
\hfil\break\indent
In general, the existence of this kind of a bound state of a $^3$He atom,  localized either at a free liquid $^4$He surface, or at the interface between crystalline and superfluid phases, is an issue of interest not only for its relevance to different aspects of the phenomenology of superfluid or solid $^4$He, but also as a pathway to the stabilization of a quasi-2D $^3$He gas with novel, fascinating superfluid properties \cite{edwards1978,bashkin1980,miyake1983}.
There is some experimental evidence of $^3$He substrate bound states, but it seems fair to define it indirect, and not entirely conclusive \cite{carmi1988,wang1992,ketola1993,draisma1994,ross1995,rolley1995,chang2021}. 
\hfil\break\indent
Only recently, over three decades after the original study of  Ref. \cite{treiner1993}, has an independent microscopic study been carried out \cite{boninsegni2022}, attempting to assess quantitatively the predictions, and possibly overcoming (some of) the limitations, of the calculation of Ref. \cite{treiner1993}, which is based on a heuristic density functional. The study of Ref. \cite{boninsegni2022} made use of first principle, Quantum Monte Carlo (QMC) computer simulations and utilized the same mathematical model of Ref. \cite{treiner1993}, focusing however on the limit of a macroscopically thick $^4$He film on a $^4$He substrate. While that calculation confirmed qualitatively some of the DFT predictions, significant quantitative differences were reported. In particular, the $^4$He density profile $n(z)$ in the direction perpendicular to the substrate computed in Ref. \cite{boninsegni2022} displays considerably more marked oscillations in the vicinity of the substrate, compared to that yielded by DFT. This can significantly affect the occurrence of a bound state of a $^3$He atom near the substrate, as one can semi-quantitatively understand based on Lekner's approach, in which $n(z)$ is related to the effective potential ``seen'' by a $^3$He atom, then used in a single-particle Schr\"odinger equation \cite{lekner1970,treiner1993}.
\hfil\break\indent
In this paper, we extend the study of Ref. \cite{boninsegni2022} and carry out a more extensive comparison of the results with those of Ref. \cite{treiner1993}. Specifically, we consider weakly attractive (alkali) substrates, this time for thin (i.e., $\lesssim 10$ layers) $^4$He films. The main goal of this study is to investigate the binding of a single $^3$He atom to a thin $^4$He film wetting the substrate, and specifically whether the bound state(s) are primarily residing on the free $^4$He surface, or in some cases closer to the substrate (obviously for films of sufficient thickness for which such a distinction is meaningful). Here too, we make use of essentially the same theoretical model adopted in Ref. \cite{treiner1993} (see below for details). 
\hfil\break\indent
Our results yield no evidence of $^3$He ``substrate'' states for thin $^4$He films adsorbed on the alkali substrates considered here, namely Cs and Li, which are, respectively, the most weakly and strongly attractive; we have also carried out a few calculations for a Na substrate as well, in order to compare our results {\em directly} to those of Ref. \cite{treiner1993}. In particular, we find that in the low temperature limit, for the substrates considered here (and, by extension, all alkali substrates), and all film thicknesses, the probability density of position of the $^3$He atom is strongly peaked at the free $^4$He surface.
We find once again {\em major quantitative} differences between the computed $^4$He density profiles and those reported in Ref. \cite{treiner1993}, suggesting that the scenario of substrate bound states may well be a spurious results, a consequence of a crude approximations built into the original DFT approach. Subsequent versions thereof (see, for instance, Ref. \cite{ancilotto2005}) might afford better quantitative agreement with first principle calculations \cite{boninsegni1999}, and it may be interesting to utilize them to reassess the prediction of the existence of substrate bound states of $^3$He.
\hfil\break\indent
This paper is organized as follows: in section \ref{model} we describe the model of interest and briefly review the computational technique utilized; in section \ref{res} we present our results, outlining our conclusions in section \ref{concl}.
\section{Model and Methodology} \label{model}
The model system simulated here comprises $N$ He atoms, one of them being of the light isotope $^3$He and all others of $^4$He; all the atoms are assumed to be point-like. The system is enclosed in a parallelepipedal cell  of sizes $L\times L\times L_z$. The substrate is modeled as a flat surface (i.e., corrugation is neglected), located at $z=0$; periodic boundary conditions are used in all three directions, but the length $L_z$ of the cell along $z$  is taken long enough to make boundary conditions irrelevant in that direction. The nominal helium coverage is $\theta=N/L^2$. 
\hfil\break\indent
The interaction between two He atoms is described by the accepted Aziz pair potential \cite{aziz1979}, while that of the He atoms with the substrate is accounted for by means of a potential that only depends on the distance of the atom from the substrate. In most of our calculations, we made use of the atom-substrate potential of Chizmeshya {\em et al.} (CCZ) \cite{chizmeshya1998}, but we also performed a few calculations with the simpler, widely adopted ``3-9'' potential, in order to compare our results directly to those of Ref. \cite{treiner1993}, in which such a potential was used. The main difference between the two potentials is that, for a given alkali substrate, the CCZ model interaction has a significantly deeper attractive well, yielding values of wetting temperature and contact angle in closer agreement with experiment.
The most important approximation contained in this model, one that is invariably made when investigating helium adsorption on weak substrates, is the neglect of substrate corrugation, which seems justified on account of the relatively large distance ($\gtrsim 4$ \AA) at which the atoms on the first adsorbed layer sit, even for the most attractive alkali substrate, namely Li \cite{boninsegni1999}.
\hfil\break\indent
The QMC methodology adopted here is the canonical \cite{mezz,mezz2} continuous-space Worm Algorithm \cite{worm,worm2}, a finite temperature ($T$) quantum Monte Carlo (QMC) technique. Although we are clearly interested in low temperature ($T\to 0$) physics, 
finite temperature  methods have been repeatedly demonstrated to have several advantages over ground state ones, for investigating Bose systems (for an extensive discussion  of this subject, see for instance Ref. \cite{rmp95}); in particular, they are unaffected by the bias of existing ground state methods (e.g., Diffusion Monte Carlo), arising from the use of a trial wave function, as well as from the control of a population of walkers \cite{psb,phasesep}, often leading to incorrect predictions (see, for instance, Refs. \cite{boninsegni2013,cinti2017,cinti2019}). The results presented in this work were obtained by carrying out simulations at temperature $T=0.25$ K, which we empirically found to be low enough that the physical estimates obtained at this temperature can be regarded as ground state estimates. 
\\ \indent
Other details of the simulations carried out in this work are standard, and therefore the reader is referred to the original references. We used the fourth-order approximation for the short-time imaginary-time propagator (see, for instance, Ref. \cite{jltp2}), and observed convergence of all the physical estimates for a value of time step $\tau=3.125\times 10^{-3}$ K$^{-1}$.
\\ \indent
The quantities of interest in this work are mainly structural, namely the integrated $^4$He density profile $n(z)$ computed along the direction perpendicular to the substrate, as well as the probability density of position of the lone $^3$He atom, used to assess the nature of the bound state.  Most of the results shown here were obtained by simulating systems comprising up to 242 atoms (this for the largest coverage considered here, namely $\theta=0.6$ \AA$^{-2}$). Experience accumulated in decades of computer simulations of thin $^4$He films adsorbed on a variety of substrates, gives us reasonable confidence that the system size utilized here affords quantitatively reliable energetic and structural estimates. 
\section{Results}\label{res}
\begin{figure} [t]
\centering
\includegraphics[width=\linewidth]{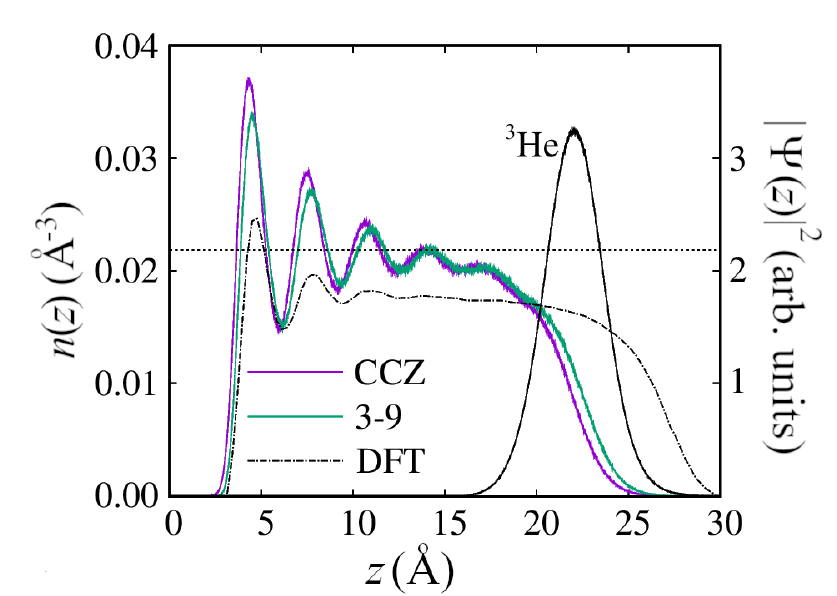}
\caption{$^4$He density profile $n(z)$ in the direction perpendicular to a Na substrate, computed in this work, for a helium film of coverage $\theta=0.4$ \AA$^{-2}$ at $T=0.25$ K. Results are shown for two different choices of the potential describing the interaction of the helium atoms with the substrate, namely the CCZ and the 3-9 (see text). Dashed line shows the DFT result from Ref. \cite{treiner1993}. Also shown (labeled $^3$He), is the probability density of position $|{\rm \Psi}(z)|^2$ of the single $^3$He atom (right axis, arbitrary units). The dotted horizontal line shows the $^4$He $T=0$ equilibrium density $n=0.021837$ \AA$^{-3}$.}
\label{f1}
\end{figure}
We now proceed with the illustration of our results, for convenience organized by substrate.
\bigskip

\noindent
{\em Sodium}. Fig. \ref{f1} shows the $^4$He density profile $n(z)$ computed by QMC for a helium film of coverage $\theta=0.4$ \AA$^{-2}$. For this particular case, we compare the results obtained with two different model potentials describing the interaction of helium atoms with the substrate, namely the CCZ and the 3-9, on which the calculation of Ref. \cite{treiner1993} is based\footnote{The parameters for this particular potential are taken from Table 1 of Ref. \cite{treiner1993}. They are different from those reported in Table 1 of Ref. \cite{chizmeshya1998}.  }. 
\\ \indent
While there are noticeable differences between the $^4$He density profiles computed with the two different potentials, chiefly higher peaks yielded by the CCZ (which is more attractive), such differences are negligible compared to the much wider discrepancy existing between the QMC and the DFT result (dashed line in Fig. \ref{f1}, the data were read off Fig. 2 of Ref. \cite{treiner1993}). In particular, besides the fact that the density peaks yielded by the DFT calculation are considerably less pronounced than those computed by QMC (the height of the main peak is lower by as much as $\sim 30$ \%), DFT predicts altogether a $^4$He film whose average density is very low (shown for comparison in Fig. \ref{f1} is the equilibrium density of superfluid $^4$He at $T=0$, dotted line), extending some $\sim 5$ \AA\ further away from the substrate, compared the QMC-computed $n(z)$, with either choice of helium-substrate potential. Also shown in Fig. \ref{f1} is the probability density of position $|{\rm\Psi}(z)|^2$ of the single $^3$He atom, which is strongly peaked in correspondence of the free $^4$He surface.
\\ \indent 
What about substrate states?  For the case of sodium, within DFT the $^3$He substrate state was found to be rather sharply localized in the proximity of the substrate, at a slightly lower distances than the first adsorbed $^4$He layer, and to have a binding energy comparable to that of the bound state located at the $^4$He surface (Fig. 2 of Ref. \cite{treiner1993}) (roughly close to 5 K). 
A substantial underestimation by the DFT approach of the local $^4$He density, which the results shown in Fig. \ref{f1} clearly  suggest, may drastically alter quantitatively the physical scenario of Ref. \cite{treiner1993}, mainly by increasing significantly the kinetic energy of localization associated to the putative substrate bound state. In any case, no evidence of any substrate bound state was found in our QMC study.
\bigskip
\begin{figure} [t]
\centering
\includegraphics[width=\linewidth]{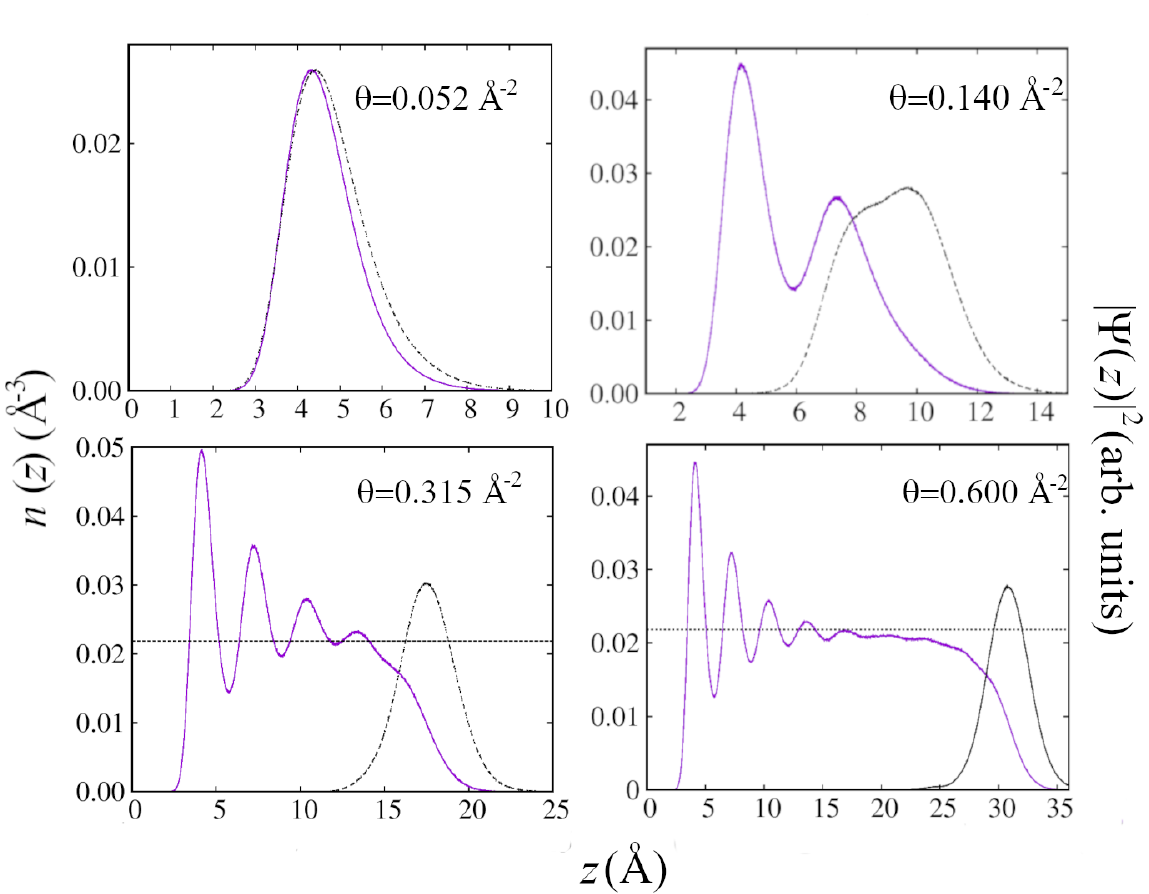}
\caption{$^4$He density profile $n(z)$ in the direction perpendicular to a Li substrate, computed in this work, for helium films of coverage $\theta$ from 0.052 to 0.600 \AA$^{-2}$ at $T=0.25$ K. Also shown (dashed lines), is the probability density of position $|{\rm \Psi}(z)|^2$ of the single $^3$He atom (right axis, arbitrary units). Horizontal lines show the $^4$He $T=0$ equilibrium density.}
\label{f2}
\end{figure}

\noindent
{\em Lithium}. This is the most strongly attractive of the alkali substrates, one on which $^4$He has been predicted to form a quasi-two-dimensional superfluid monolayer at low temperature \cite{boninsegni1999}. Fig. \ref{f2} shows $^4$He density profiles $n(z)$ computed on a Li substrate, at $T=0.25$ K, for four different coverages, from the $T=0$ equilibrium density $\theta=0.052$ \AA$^{-2}$ \cite{boninsegni2004} to 0.600 \AA$^{-2}$, together with the probability density of position of a substitutional $^3$He atom. At the equilibrium density, $^4$He forms a (superfluid) quasi-2D monolayer on a Li substrate, and, as shown in Fig. \ref{f2}, the single $^3$He atom lies very nearly on the same plane of the $^4$He atoms. As the coverage is increased, no solid layer forms, due to the weakness of the helium-substrate potential. Crystallization is preempted by promotion of atoms to the second layer, with the concurrent displacement of the $^3$He atom to the surface of the film, where it remains as successive layers form, i.e., the $^3$He atom only binds to the film near the substrate in the presence of a single $^4$He layer.
There is no evidence of a bound state of $^3$He localized near the substrate for any film thickness exceeding one layer.
\\ \indent
The $^4$He density profiles for the two largest coverages shown in Fig. \ref{f2} suggest that, in the limit of a thick film, $n(z)$ fairly quickly stabilizes to a value close to the superfluid $^4$He $T=0$ equilibrium bulk value, displaying three well-defined oscillations in a substrate region of width $\sim$ 10 \AA. If we tentatively assume that the result shown in Fig. \ref{f2} for $\theta=0.6$ \AA$^{-2}$ provides a quantitatively reliable representation of  $n(z)$ in the substrate region for the case of a thick film, then we may compare this result to that shown in Fig. 3 of Ref. \cite{treiner1993}. In this case too, just like for the Na substrate, the original DFT study yielded considerably less marked density oscillations in the substrate region; for example, the height of the first (second) peak is underestimated by $\sim$ 10 (20)\%.
\bigskip
\begin{figure} [t]
\centering
\includegraphics[width=\linewidth]{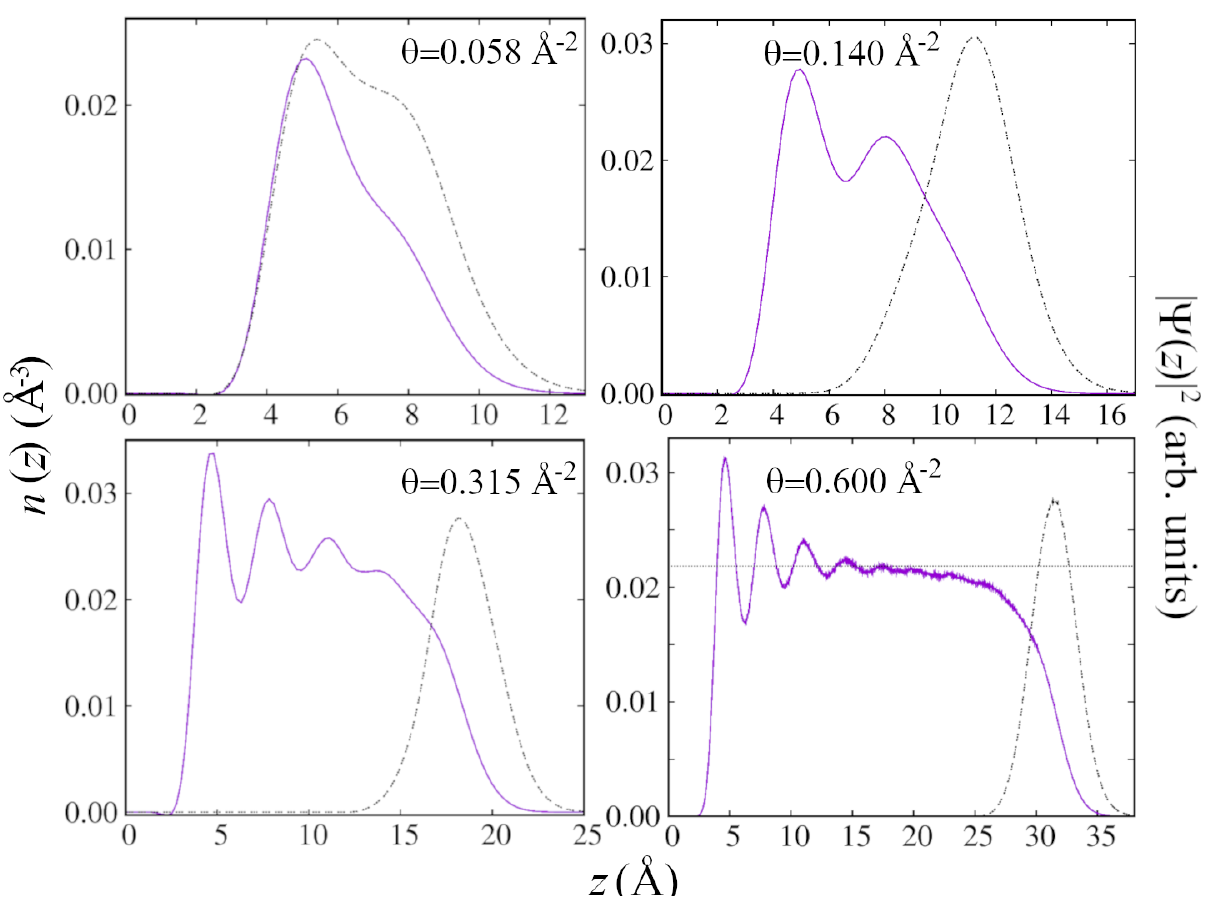}
\caption{$^4$He density profile $n(z)$ in the direction perpendicular to a Cs substrate, computed in this work, for helium films of coverage $\theta$ from 0.058 to 0.600 \AA$^{-2}$ at $T=0.25$ K. Also shown (dashed lines), is the probability density of position $|{\rm \Psi}(z)|^2$ of the single $^3$He atom (right axis, arbitrary units). Horizontal line shows the $^4$He $T=0$ equilibrium density.}
\label{f3}
\end{figure}

\noindent{\em Cesium.} On the weakest substrate, namely Cs, $^4$He is  {\em not} expected to form a thin film at low $T$, due to the shallowness of the attractive well of the helium-substrate potential \cite{cheng1991}. Indeed, experimental evidence shows that $\theta\sim0.9$ \AA$^{-2}$ (roughly eleven layers) is the minimum coverage for which a $^4$He film will form on a Cs substrate \cite{ketola1992}; for lower coverages, the film breaks down into droplets, characterized by a well-defined contact angle with the surface \cite{cheng1991,klier1995,rolley1998,ross1998}. For a thick $^4$He film on Cs, substrate states for $^3$He atoms were predicted to exist, in Ref. \cite{treiner1993}. 
\\ \indent
It was proposed in Ref. \cite{pettersen1993} that a small concentration  of $^3$He alters the above state of affairs, stabilizing thin films of helium on Cs at low $T$. This prediction was verified experimentally for helium mixtures with a $^3$He concentration of a few percent \cite{ketola1993,ross1995}. Based on this prediction and its observation, we have carried out simulations of thin $^4$He films on a Cs substrate, assuming a low $^3$He concentration, sufficient to stabilize the film\footnote{It should be noted that the methodology utilized in this work allows one to observe, in a computer simulation, a thin film of pure $^4$He on a Cs substrate ``bead up'' and form a single droplet, at the temperature considered in this study. However, this requires that systems of sufficiently large size be studied. For $N \lesssim 200$ and/or at low coverage, periodic boundary conditions stabilize a uniform thin film.}.
\\ \indent
Fig. \ref{f3} shows the results of our computer simulations for a Cs substrate. There is little qualitative difference with respect to the results of Fig. \ref{f2}, i.e., for a Li substrate. If anything, the tendency for the $^3$He atom to be pushed to the surface appears {\em accentuated}, compared to what observed on a Li substrate. And in this case too, the comparison of the result for $n(z)$ at the highest coverage considered, namely $\theta=0.6$ \AA$^{-2}$, with those of Ref. \cite{treiner1993} for a thick film (Fig. 3 therein), points to a significant underestimation (as much as 30\%) by the DFT calculation of the amplitude of the local density oscillations in the vicinity of the substrate. 
\\ \indent
It is worth pointing out the the results for the four coverages shown in Fig. \ref{f2} and Fig. \ref{f3} are only representative; results for many other intermediate coverages are available, and none of them yield any evidence of a competition between the surface and the substrate, in terms of where the $^3$He impurity may reside.

\section{Discussion and Conclusions}\label{concl}
First principle numerical studies of adsorption of $^4$He films on alkali substrates at low temperature yield no evidence of the substrate bound states of $^3$He atoms predicted almost thirty decades ago, by a DFT calculation. These calculations are based on the same microscopic Hamiltonian of which the original DFT calculations made use. Comparison of the $^4$He density profiles in the direction perpendicular to the substrate shows that the DFT study quantitatively underestimated the local $^4$He density oscillations in the vicinity of the substrate. \\ \indent As shown by Lekner's theory, the prediction of a possible $^3$He bound state near the substrate hinges on a rather delicate energy balance, which heavily relies on an accurate representation of the local $^4$He density near the substrate. To our knowledge no subsequent study of this problem has been carried out, based on DFT, following that of Ref. \cite{treiner1993}; it seems conceivable that the original DFT prediction may have to be revised, possibly with the aid of an improved density functional. Indeed, much more satisfactory agreement between QMC and DFt was eported in subsequent works \cite{boninsegni1999}.
\\ \indent
At the same time, it must also be noted that some experimental evidence that can be interpreted as supporting the existence of $^3$He substrate states has been reported \cite{ketola1993,ross1995}. While that evidence may not be entirely conclusive, nonetheless it seems fair to conclude that the issue of the existence of substrate states is to some extent still open.
 
\section*{Acknowledgments}
This work was supported by the Natural Sciences and Engineering Research Council of Canada. Computing support of ComputeCanada is acknowledged.

\section*{Conflict of interest}
The author declares no conflict of interest.

\bibliographystyle{spphys}       
\bibliography{refs}   

\end{document}